\documentclass[12pt]{iopart}
\usepackage{colordvi,amssymb,color}
\usepackage[dvips]{graphicx}

\begin{document}
\title[Non-linear growth model]{A continuous non-linear shadowing  model of columnar growth}
\author{T.H. Vo Thi$^1$, 
J.-L. Rouet$^{1,2}$, 
P. Brault$^3$ \footnote[3]{
Corresponding author: Fax +33(0)2 38 41 71 54, e-mail: 
\mailto{Pascal.Brault@univ-orleans.fr}},
J.-M. Bauchire$^3$, 
S. Cordier$^1$
 and C. Josserand$^4$}
\address{$^1$ Laboratoire de Math\'ematiques et Applications, Physique
Math\'ematique d'Orl\'eans, UMR6628
CNRS-Universit\'e d'Orl\'eans BP 6759, 45067 Orl\'eans Cedex 2, France }
\address{$^2$ Institut des Sciences de la Terre d'Orl\'eans
UMR6113 CNRS/Universit\'e d'Orl\'eans, 1A rue de la F\'erollerie, 45071 ORLEANS
CEDEX 2}
\address{$^3$ Groupe de Recherches sur l'Energ\'etique des Milieux Ionis\'es,
UMR6606 CNRS-Universit\'e d'Orl\'eans BP 6744, 45067 Orl\'eans Cedex 2, France}
\address{$^4$ Institut Jean Le Rond d'Alembert, UMR 7190 CNRS-Paris VI Case 162, UPMC, 4 place Jussieu, 75252 Paris Cedex 05,  France }
\eads{\mailto{Jean-Louis.Rouet@univ-orleans.fr},
\mailto{Stephane.Cordier@univ-orleans.fr}}

%
\begin{abstract}
We propose the first continuous model with long range screening (shadowing) that described columnar growth in one space dimension, as observed in plasma sputter deposition. It is based on a new continuous partial derivative equation with non-linear diffusion and where the shadowing effects 
apply on all the different processes.
\end{abstract}
\pacs{81.15.Aa, 68.35.Ct, 05.40.-a}
\submitto{\JPD - Fast Track Communications}
\maketitle
%

Plasma sputtering is a common process for film growth which often exhibits wide columns more or less close packed separated by thin deep grooves \cite{BZ,RaBr08,messier,dirks}. This columnar growth mainly results  from a shadowing instability\cite{bales,karu,roland}, where the elevated parts of the 
surface are more exposed to the sputtering while they shadow the incoming particles to the lower parts. 
The modelization of this shadowing instability has been well described by probabilistic 
Monte-Carlo methods (MC)\cite{tombo,meng,ficht} and also with continuous models based 
on partial derivative equations (PDE)\cite{karu,roland,sarma,rost,amar,yao1,yao,drotar} including
the seminal work of Bales and Zangwill \cite{BZ}. 
However, both approach fail 
to describe at long times the strongly nonlinear columnar microstructures observed recently (see \cite{RaBr08} for instance). In fact, although the continuous models gives tall and well separated columns at early time,
 only few sharp peaks remain later on\cite{karu,yao1,drotar}. Columnar structure using PDE has
 already been obtained by Gillet {et al.}\cite{gillet} but in that case no 
shadowing effect was taken into account! On the other hand, discrete approaches using MC methods 
including shadowing have been developped and showed a fair description of the  columnar structure,
particularly through the formation of sharp column sides. However, these models cannot avoid
the coarsening of the columnar structures showing larger and larger plateau as time
increases, in contrast with experimental observations.

\begin{figure}[!h]
\centerline{\includegraphics*[width=.3\linewidth]{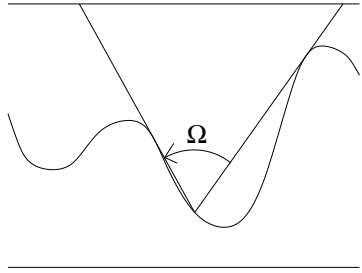}}
\caption{\label{MC2-1} \it Shadowing process : interface $h(x,t)$ and solid angle $\Omega(x,\{h\})$.}
\end{figure}

The goal of this paper is to present a new continuous non-local model which includes both non-linear shadowing and diffusion effects to simulate columnar-like growth. We consider a two dimensional model
where the one dimensional (1D) surface described by $h(x,t)$ is subjected to receive particles from all
directions not shadowed by the surface itself. Our starting point is deduced from the models developped
initially by Bales and Zangwill\cite{BZ} and by Karunasiri et al \cite{karu}:

\begin{equation}
\frac{\partial h}{\partial t}=R\Omega(x,\{h\})\sqrt{1+(\nabla h)^2} +\nu \nabla^2 h +  \eta
\label{eq1.0}
\end{equation}

where the deterministic deposition term $R$ is
multiplied by the solid angle $\Omega(x,\{h\})$ which modelizes the
shadowing effect as a long range screening (see figure \ref{MC2-1}).  $\nu$ is the 
diffusion/relaxation coefficient while $\eta$ is the usual noise with zero mean $<\eta> = 0 $ and 
its correlation given by $<\eta(x,t)\eta(x',t')>=2D\delta(x,x')\delta(t,t')$.\\

For small surface angles, we retrieve a KPZ-like equation \cite{kardar} with shadowing effects
(defining $\lambda=\pi R$):

\begin{equation}
\frac{\partial h}{\partial t}=\nu \nabla^2 h+\frac{\lambda}{2}(\nabla h)^2 +
R\Omega(x,\{h\}) + \eta.
\label{eq1.1}
\end{equation}

A complete study of these equations has shown that it is unable to reproduce columnar shapes corresponding to experiments and MC simulations \cite{roland,yao1}. Indeed, in the most favourable situation, only broad peaks emerge instead of flat columns. Experiments suggest thus that the 
diffusion should be enhanced in the region more exposed to the flux. Moreover, we will assume that
the flux also increases (greater than for normal shadowing) on the top of the columns 
compared to the grooves. Although we have no strong argument for it, we expect some
point effect near the sharp edge to be responsible of this process.
We then propose the following stochastic differential equation where
the main ingredients are non-linear shadowing effects and
diffusion~:

\begin{equation}
\frac{\partial h}{\partial t}=g(\Omega(x,\{h\}))\,(R\,\sqrt{1+|\nabla(h)|^2}+\nu
\nabla^2 h+ \eta)
\label{eq2.0}
\end{equation}

In this equation $g(\Omega)$ is a given function of the solid angle $\Omega$. Therefore, in order to 
increase the shadowing effect and the diffusion from top to edges and bottom, $g(\Omega)$ has to be stronger than linear, and we will take later on for the numerics $g(\Omega)=\Omega^2$ . 
The fact that this function, which modelizes the
shadowing,  is in factor to the right hand term  
will obviously increase the deposit rate for surfaces which are not shadowed (mainly
for large value of $h$) and make it smaller for shadowed one (for small value of
$h$). The diffusion is also affected by the shadowing in the same way. \\

\begin{figure}[!h]
\includegraphics*[angle=-90,width=0.6\linewidth]{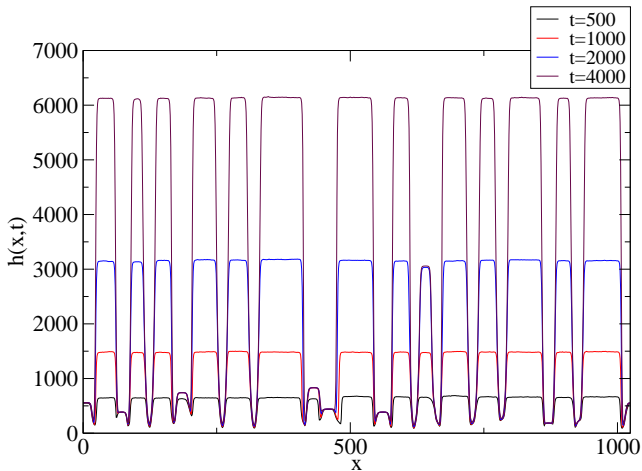}
\caption{\label{shadow1} Continuous model. Snapshots of the interface given by the
non-linear shadowing anisotropic
 diffusion model given by equation (\ref{eq2.0}) at time $t=500$, $1\,000$, $2\,000$
and $4\,000$. The numerical simulation was done with $\Delta t=0.01$, $\Delta x=1$ and the total suze of the system is $L=1024$.}
\end{figure}

A plane-wave analysis performed on equation (\ref{eq2.0}) shows that the solutions are unstable for 
large enough wavelengths $\lambda$, i.e. $\lambda=2\pi/k > \lambda^*=\nu \pi^3/(\alpha R)$,  with $\alpha  \sim 0.724 $ and the growth rate $\sigma=k \pi (2 \alpha R-\nu \pi^2 k)$. Then, starting from a flat substrate,  the noise will trigger the
instability and will drive the system into a strong non-linear regime.
Figure \ref{shadow1} shows  the evolution of the interface profile for
different times for $D=1$, $\nu=1$ and $R=1$. It  exhibits the desired 
columnar  shape. This shape is characterised by flat column tops and vertical sides as 
compared to previous Monte-Carlo simulations and columnar growth experiments.  The shadowed deposition 
favors the columnar growth and  the anisotropic diffusion smoothes the top of the column very 
efficiently and leads to vertical sides.
The competition between these two effects leads to a columnar regime as expected.
Moreover, most of the columns formed at the beginning of the simulation are still present at the end. 
This is also the case for ``Poisson/Wedding cakes'' morphologies for which initial columns 
always remains \cite{jkrug} and for step meander process \cite{gillet}.\\

For numerical simulation, equation (\ref{eq2.0}) is integrated using the following explicit scheme:

\begin{eqnarray}
h_i^{n+1} & = & h_i^n
                +\left(\Omega_i^n\right)^2\,\\ \nonumber
                & & \left[\Delta
t\,R\,\sqrt{1+\frac{(h_{i+1}^n-h_i^n)^2+(h_{i+1}^n-h_i^n)(h_i^n-h_{i-1}^n)+(h_i^n-h_{i-1}^n)^2}{3\,\Delta
x^2}}\right.\\ \nonumber
              &   &  
                + \frac{\nu\,\Delta t}{\Delta x^2}\,(h_{i+1}^n-2\,h_i^n+h_{i-1}^n)
                +\left. \sqrt{\frac{2\,D\,\Delta t}{\Delta x}}\,\,\varepsilon\right] 
\label{eq3.0}
\end{eqnarray}
with the notation $h_i^n=h(i\,\Delta x, n\,\Delta t)$. $\varepsilon$ is a random
number picked with the uniform distribution between $[-1,1[$. To  obtain a discrete form of the 
gradient term, we follow the scheme proposed by Lim {\it et al.} although their study strictly 
applied for the KPZ equation \cite{lam}. $\Omega(x,\{h\})$ is evaluated following reference \cite
{karu}.

\begin{figure}[!h]
\includegraphics*[angle=-90,width=0.6\linewidth]{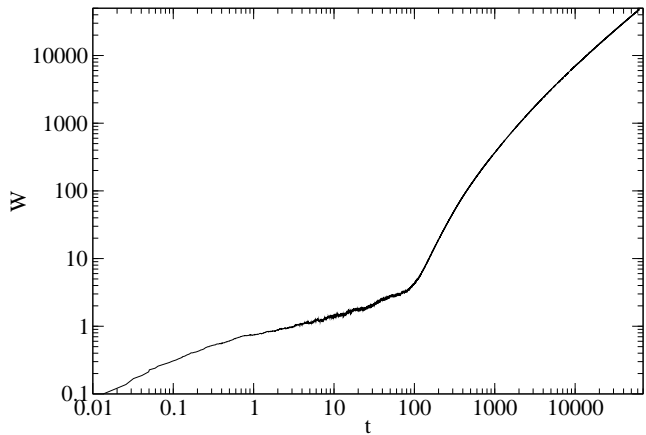}
\caption{\label{shadow3} \it Continuous model. Roughness $W(t)$ as a function of time.}
\end{figure}

The time evolution of the roughness $W$ of the interface is given figure
\ref{shadow3}. It shows the existence of different regimes. The
first one, for $t<1$ is driven by the fluctuations and $W$ scales as $t^{0.5}$. For
the second one ($1<t<100$), diffusion induced a relative reduction of the roughness
which scales as $t^{0.4}$. Then, because of the shadowing instability described above, sharp canyons 
appear and the roughness quickly increases. Finally, after $t\sim 1000$, the columnar regime appears 
which leads to $W(t)\sim t$ as in the discrete model \cite{yao1}. Even if $W(t)$ shows the same scaling 
as obtained in previous studies on continuous columnar growth model \cite{yao1, yao, drotar}, the 
column shapes are rather different and are now in closer agreement with the MC models and 
more important with the
experiments \cite{messier,dirks}. Indeed, experiments display relative constant column width, while 
MC simulations lead to a column width which is increasing with time. In
that respect, our model (\ref{eq2.0}) exhibits a better qualitative agreement with experiement than MC simulations.

We have presented the first continuous model, to our knowledge, that exhibits a columnar
growth without the coarsening dynamics of the structures, in good agreement with experimental
observations on sputtering deposition. For reproducing these wide flat columns with sharp edges,
 we have introduced 
an increase of the relaxation and  of the flux on the top of the column compared to
the grooves. By sake of simplicity, we have considered a 1D surface and we have taken the
same non local and nonlinear multiplicative factor  $g(\Omega(x,\{h\}))$ for all the terms of the dynamics.
Further on, we have considered a simple power law behaviour for this function $g(\Omega(x,\{h\})) = \Omega^n$. We argue that $n>1$ is needed to enhance the shadowing effects on the protuberances. 
We have tested numerically $n=2$ and $n=3$ with no loss of properties of the results. However, a
better choice of the shadowing function $g$ should be obtained through further experimental
comparisons. Similarly, different shadowing functions should be considered in the future for the 
diffusion term and the deposition term.
Finally, this new continuous model, with $n=2$, considered as a minimal model, already 
correctly reproduces the formation of flat wide columns, with sharp edges and thin separating 
grooves, as usually encountered in sputtering deposition. Further works should perform such 
approaches to two dimensional surfaces in particular.\\

\section*{Acknowledgments}
French Ministry of Research, Conseil R\'egional du Centre, Communaut\'e d'Agglom\'eration du Drouais, Conseil G\'{e}n\'{e}ral de l'Eure et Loir  and CRT Plasma-Laser are acknowledged for a research grant.
C.J. wants to thank the financial suport of the DGA.

\end{document}